\documentclass[aps,prx,11pt,superscriptaddress]{revtex4-1}
\usepackage{amsmath}
\usepackage{amsthm}
\usepackage[colorlinks=true,urlcolor=blue,citecolor=blue,linkcolor=black]{hyperref}
\usepackage[shortlabels]{enumitem}
\usepackage{amsfonts}
\usepackage{graphicx}
\usepackage{bbold}
\usepackage{stmaryrd}
\usepackage{color}
\usepackage{hyperref}
\usepackage{verbatim}
\usepackage{cases}
\usepackage{amsfonts}
\usepackage{amssymb}
\usepackage[]{mathrsfs}
\usepackage{xcolor}
\usepackage{anysize}
\usepackage{epsfig}
\usepackage{bm}
\usepackage{setspace}
\usepackage{enumerate}
\usepackage{dcolumn}
\usepackage{bm}
\usepackage{enumitem}
\newcommand{\bra}[1]{\left< #1 \right\vert}
\newcommand{\ket}[1]{\left\vert #1 \right>}

\newcommand{\im}{\mathrm{i}}
\newcommand{\p}{\psi}
\newcommand{\PP}{\Psi}
\newcommand{\g}{\gamma}
\newcommand{\cc}{\kappa}

\renewcommand{\epsilon}{\varepsilon}

\usepackage[bottom]{footmisc}

\usepackage{lineno}

\begin{document}

\title{Endurance of quantum coherence due to particle indistinguishability \\ in noisy quantum networks}


\author{Armando Perez-Leija}\email{apleija@gmail.com}
\affiliation{Max-Born-Institut, Max-Born-Stra{\ss}e 2A, 12489 Berlin, Germany}
\affiliation{Humboldt-Universit\"at zu Berlin, Institut f\"{u}r Physik, AG Theoretische Optik \& Photonik, Newtonstra{\ss}e 15, 12489 Berlin, Germany}
\author{Diego Guzm\'{a}n-Silva}
\affiliation{Institute of Applied Physics, Friedrich-Schiller Universit\"at, Max-Wien-Platz 1, 07743 Jena, Germany}
\affiliation{Institut f\"ur Physik, Universit\"at Rostock, D-18051 Rostock, Germany}
\author{Roberto~de~J.~Le\'{o}n-Montiel}\email{roberto.leon@nucleares.unam.mx}
\affiliation{Instituto de Ciencias Nucleares, Universidad Nacional Aut\'onoma de M\'exico, 70-543, 04510 Cd. Mx., M\'exico}
\author{Markus~Gr\"{a}fe}
\affiliation{Institute of Applied Physics, Friedrich-Schiller Universit\"at, Max-Wien-Platz 1, 07743 Jena, Germany}
\author{Matthias Heinrich}
\affiliation{Institut f\"ur Physik, Universit\"at Rostock, D-18051 Rostock, Germany}
\author{Hector Moya-Cessa}
\affiliation{Instituto Nacional de Astrof\'{i}sica, \'{O}ptica y Electr\'{o}nica, Calle Luis Enrique Erro 1, Santa Mar\'{i}a
Tonantzintla, Puebla CP 72840, M\'{e}xico}
\author{Kurt Busch}
\affiliation{Max-Born-Institut, Max-Born-Stra{\ss}e 2A, 12489 Berlin, Germany}
\affiliation{Humboldt-Universit\"at zu Berlin, Institut f\"{u}r Physik, AG Theoretische Optik \& Photonik, Newtonstra{\ss}e 15, 12489 Berlin, Germany}
\author{Alexander Szameit}\email{alexander.szameit@uni-rostock.de}
\affiliation{Institut f\"ur Physik, Universit\"at Rostock, D-18051 Rostock, Germany}

\begin{abstract}
\begin{center}
\textbf{ABSTRACT} 
\end{center}
Quantum coherence, the physical property underlying fundamental phenomena such as multi-particle interference and entanglement, has emerged as a valuable resource upon which modern technologies are founded. In general, the most prominent adversary of quantum coherence is noise arising from the interaction of the associated dynamical system with its environment. Under certain conditions, however, the existence of noise may drive quantum and classical systems to endure intriguing nontrivial effects. In this vein, here we demonstrate, both theoretically and experimentally, that when two indistinguishable non-interacting particles co-propagate through quantum networks affected by non-dissipative noise, the system always evolves into a steady state in which coherences accounting for particle indistinguishabilty perpetually prevail. Furthermore, we show that the same steady state with surviving quantum coherences is reached even when the initial state exhibits classical correlations.
\end{abstract}
\maketitle
\begin{figure}[h!]
\centering
\includegraphics[width=14cm]{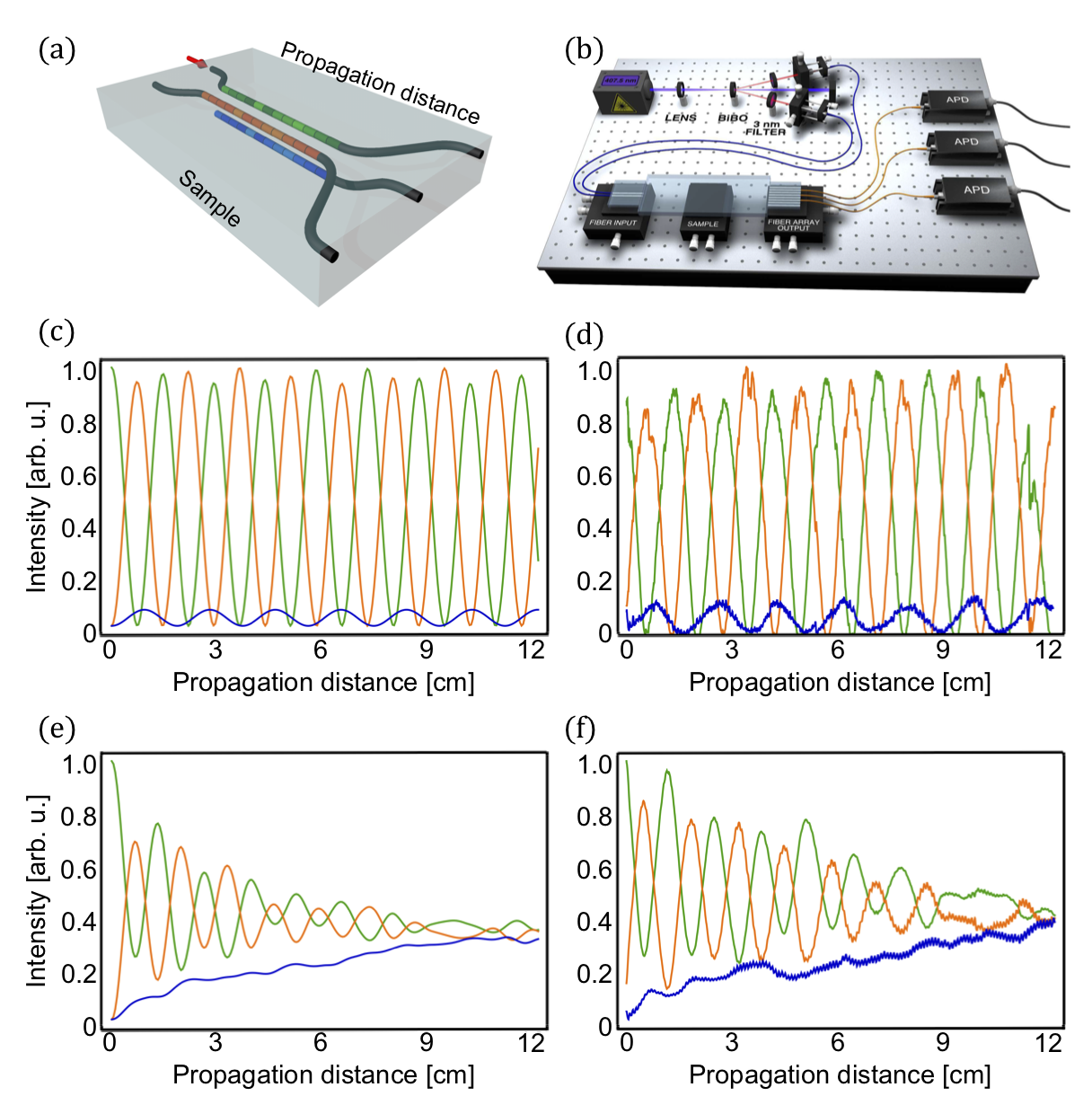}
\caption{(a) Schematic setup of an integrated waveguide trimer simulator of a reduced Born-Markov open quantum system. The different colors in the waveguides represent dynamical ($z$ dependent or time dependent) random changes in the propagation constants, whose effects emulate site-energy fluctuations induced by the environment. Notice the orange and green waveguides represent the strongly coupled sites and the blue waveguide represents the third site that is weakly coupled to the upper sites. (b) Experimental setup employed to carry out experiments within the two-excitation manifold: a two-photon source at a wavelength of 815 nm was implemented by means of spontaneous parametric down-conversion from a pump laser at a wavelength of 407.5nm. Photons emerging at the output of the device are collected via a fiber array and subsequently fed into avalanche photodiodes (APDs). In the absence of noise all waveguides have the same propagation constants. Consequently, exciting one of the uppers sites with laser light ($\lambda=633nm$) creates the intensity dynamics shown in (d). We observe that light propagates through the system hopping predominantly among the upper waveguides, i.e., the strongly coupled sites. In the presence of dephasing (f), we observe the emergence of a uniform redistribution of energy among all the waveguides. Notice in (c-f) the green curve represents the intensity along the excited waveguide, the orange curve describes the intensity along the second upper site and the blue one is the intensity recorded from the lower site, see (a). These experiments unequivocally demonstrate that within the single-excitation manifold dephasing induces a uniform redistribution of energy and as a result the farthest waveguide becomes populated with about 1/3 of the total energy. (c ,e) depict the theoretically calculated single-excitation dynamics in a noiseless and a dephasing trimer, respectively. 
}
\label{fig:F1}
\end{figure}
\begin{figure}[h!]
\centering
\includegraphics[width=14cm]{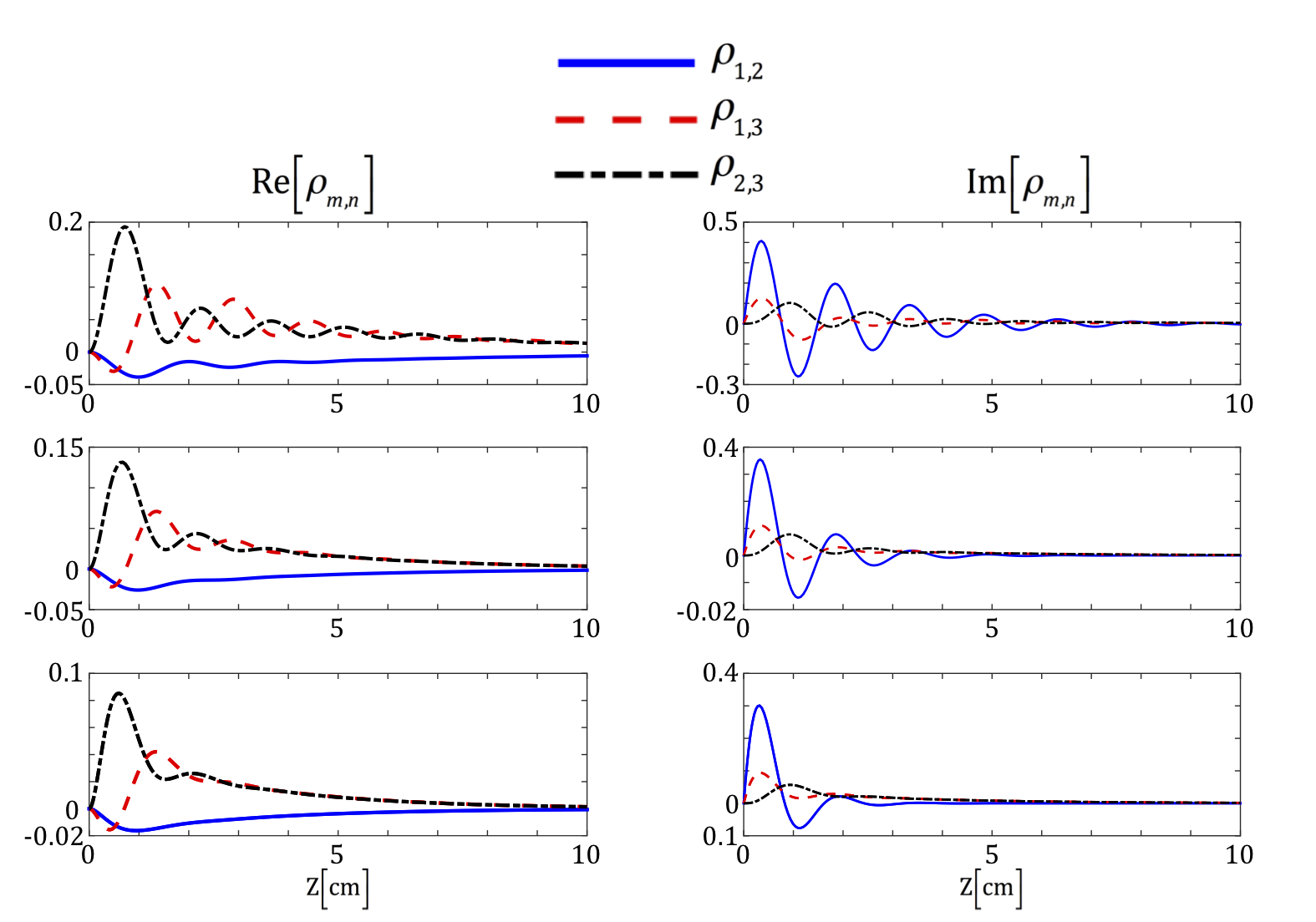}
\caption{Theoretically computed evolution of the coherence terms (off-diagonal terms $\rho_{m,n}$, $m\neq n$) of the reduced density matrices arising in waveguide trimers exhibiting dephasing rates (top) $\g=0.3\g_{exp}$, (center) $\g=0.6\g_{exp}$, and (bottom) $\g=\g_{exp}$, where $\g_{exp}=\left(\g_{1},\g_{2},\g_{3}\right)=\g_{exp}=\left(1.7275,1.7435,1.7645\right)$cm$^{-1}$ are the dephasing rates obtained from the experimental parameters used to fabricate the structures utilized in the experiments shown Fig. (\ref{fig:F1}  f). In all cases the coherences $\rho_{m,n}$ inherently vanish after certain propagation distance. The length of the simulated systems was 10cm and the propagation constants were taken from a Gaussian distribution with identical variance $\sigma=3$cm$^{-1}$, and mean values $\beta=1$cm$^{-1}$  and $\beta=-1$cm$^{-1}$ for the upper and lower waveguides, respectively. }
\label{fig:F2}
\end{figure}
\begin{figure}[h!]
\centering
\includegraphics[width=15cm]{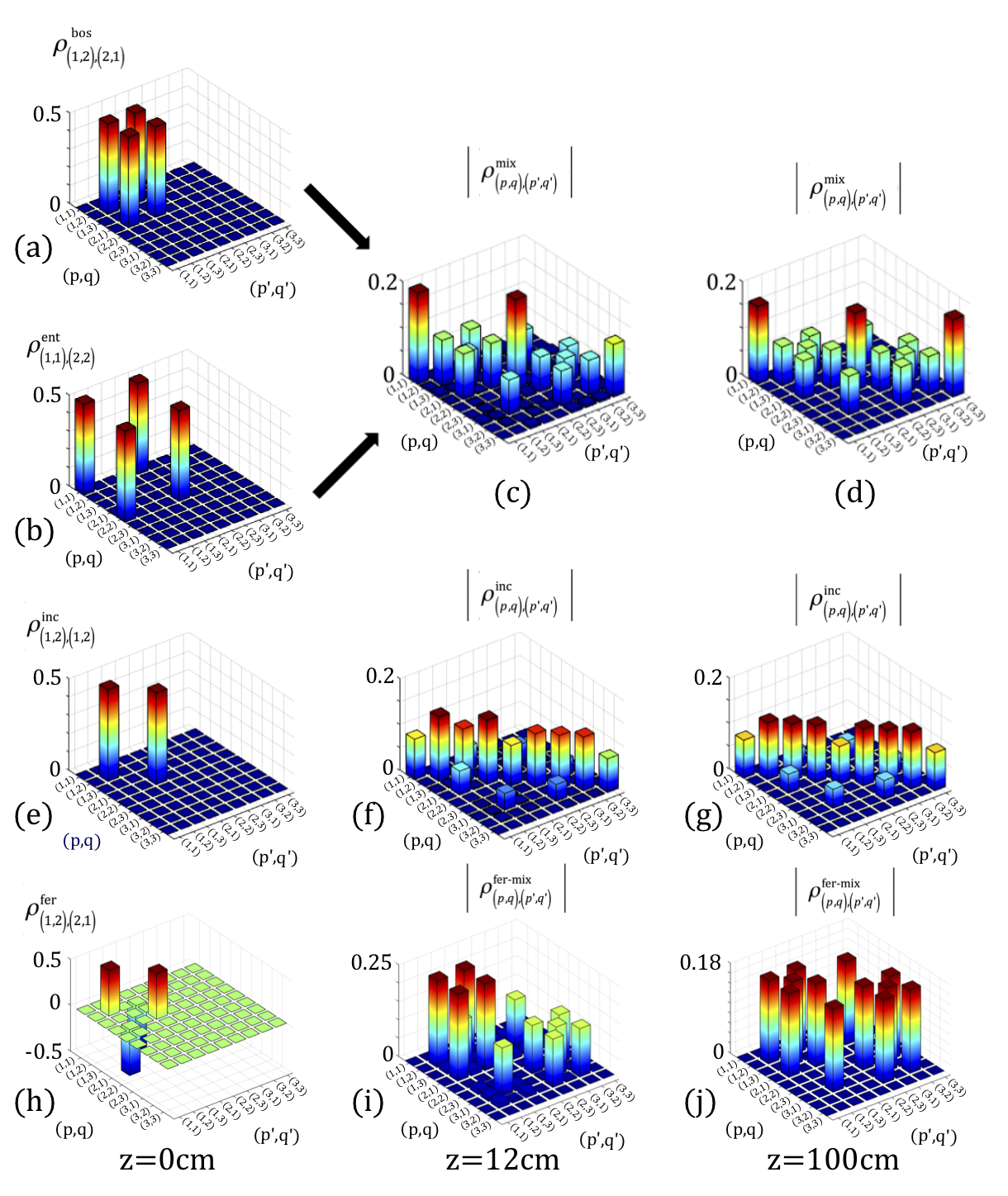}
\caption{Theoretically computed evolution of reduced density matrices for (a) separable 
$\ket{\PP^{sep}}=\frac{1}{\sqrt{2}}\left(\ket{1_{1},1_{2}}\pm\ket{1_{2},1_{1}}\right)\rightarrow \rho_{(1,2),(2,1)}=\ket{\PP^{sep}}\bra{\PP^{sep}}$,  (b) entangled 
$\ket{\PP^{ent}}=\frac{1}{\sqrt{2}}\left(\ket{1_{1},1_{1}}+\ket{1_{2},1_{2}}\right)\rightarrow \rho_{(1,1),(2,2)}^{ent}=\ket{\PP^{ent}}\bra{\PP^{ent}}$, and  (e) incoherent 
 $\rho_{(1,2),(2,1)}^{inc}=\frac{1}{2}\left(\ket{1_{1},1_{2}}\bra{1_{1},1_{2}}+\ket{1_{2},1_{1}}\bra{1_{2},1_{1}}\right)$ bosons propagating in the noisy trimer shown in Fig.~(\ref{fig:F1} a). The simulated dephasing rates correspond to the experimental ones $\g_{exp}=\left(\g_{1},\g_{2},\g_{3}\right)=\left(1.3012,1.2365,1.293\right)$cm$^{-1}$. From (c) and (d) it is clear that at $z=$12cm, separable and entangled bosons are described by identical density matrices. Once in the steady state, e.g. at $z$=100cm, the density matrices exhibit three main peaks along the diagonal, indicating that particle bunching is the most probable outcome to occur (d). In contrast, incoherent bosons exhibit a different behaviour where particle antibunching exhibits the highest probability as elucidated by the density matrices at $z$=100cm (g). In (h-j) we depict density matrices for indistinguishable fermion pairs 
$\ket{\PP^{fer}}=\frac{1}{\sqrt{2}}\left(\ket{1_{1},1_{2}}-\ket{1_{2},1_{1}}\right)\rightarrow \rho_{(1,2),(2,1)}^{fer}=\ket{1_{1},1_{2}}\bra{1_{1},1_{2}}-\ket{1_{1},1_{2}}\bra{1_{2},1_{1}}-\ket{1_{2},1_{1}}\bra{1_{1},1_{2}}+\ket{1_{2},1_{1}}\bra{1_{2},1_{1}}$. From this numerical results we see that in agreement with the Pauli exclusion principle, both fermions never occupy the same site (fermion antibunching) and the steady state contains off-diagonal entries demonstrating that some coherences survive the impact of dephasing.}
\label{fig:F3}
\end{figure}
\begin{figure}[h!]
\centering
\includegraphics[width=14cm]{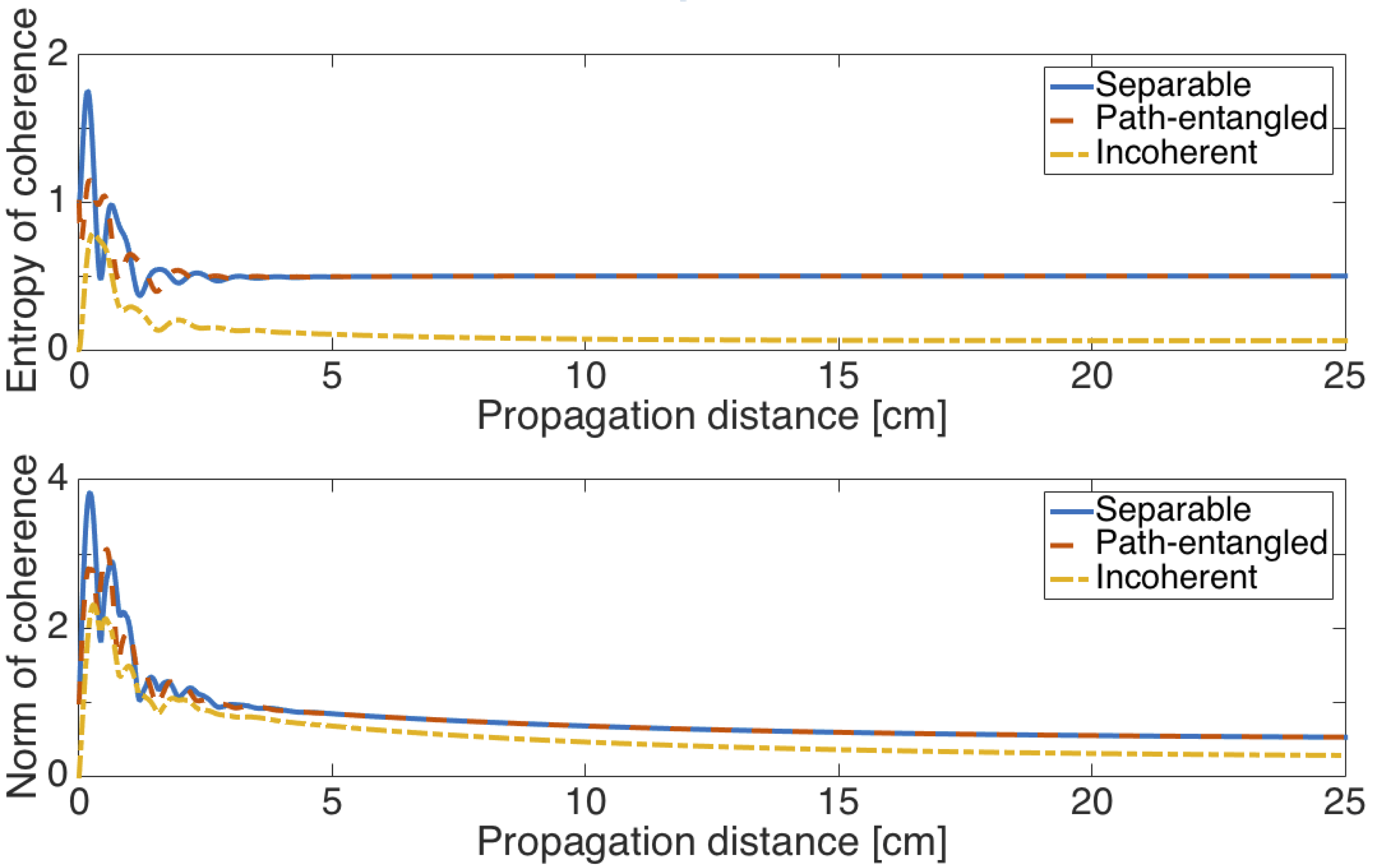}
\caption{Evolution of the relative entropy of coherence $C_{RE}(\rho)$ (top panel) and  the norm of coherence $C_{n}(\rho)$ (bottom panel) for the initial states  separable (solid line), path-entangled (dashed line), and incoherent (dash-dotted line).}
\label{fig:entropy}
\end{figure}
\begin{figure}[h!]
\centering
\includegraphics[width=14cm]{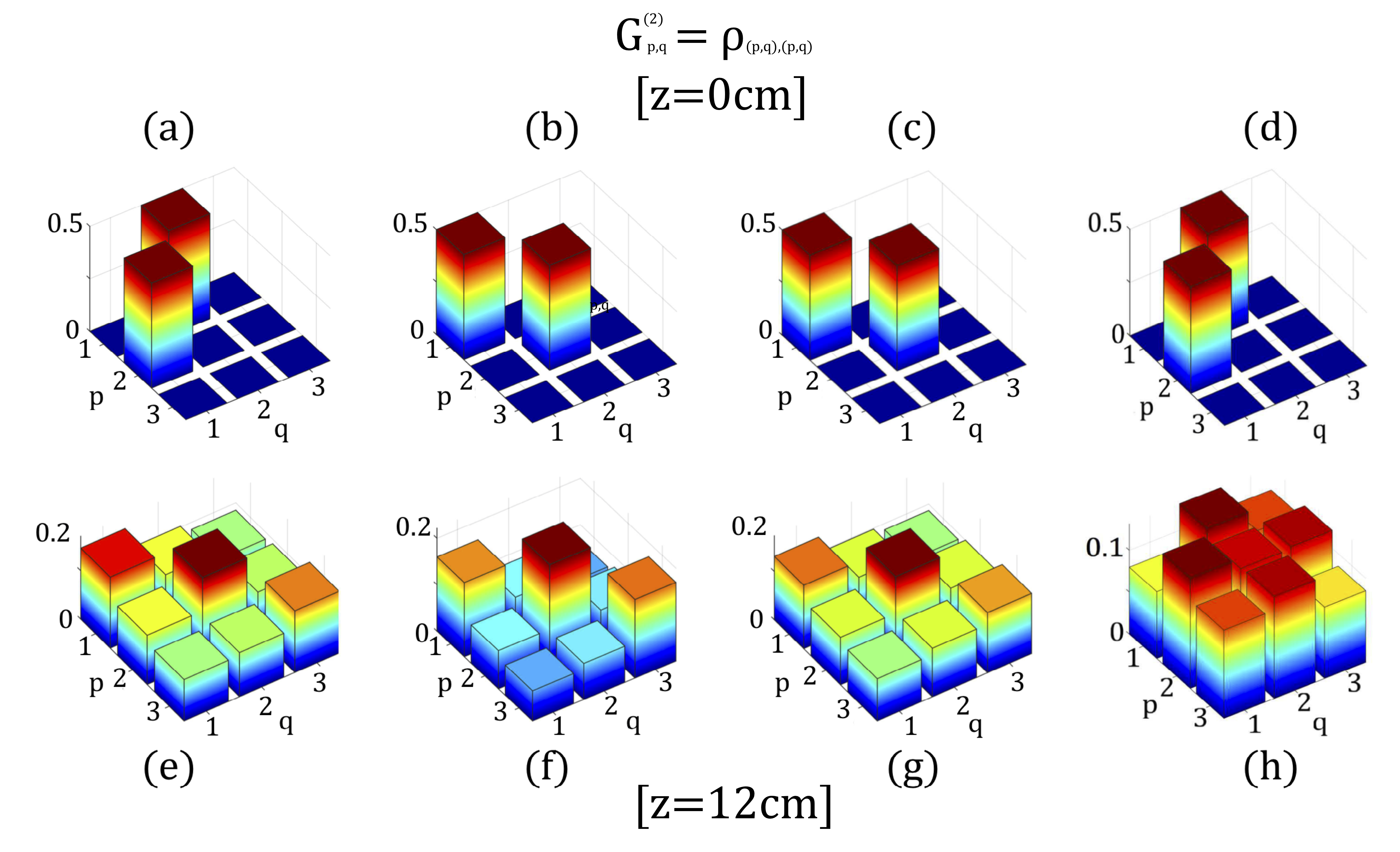}
\caption{Experimental two-point (intensity) correlation functions, $G_{p,q}^{(2)}=\rho_{(p,q),(p,q)}$, for separable (a), entangled (b), classically correlated (c), and incoherent photon pairs (d) coupled into the upper sites of the waveguide trimer shown in Fig.~(\ref{fig:F1} a). The initial states are shown in (a), (b), (c), and (d). The corresponding output correlation patterns, after a propagation distance of $z$=12cm, are shown in (e), (f), (g), and (h). From these experimental results it becomes evident that indistinguishable two-photon states produce similar correlation patterns where photon bunching exhibits the highest probability, while for incoherent distinguishable photons exhibit the highest probability in the antibunching terms.}
\label{fig:F4}
\end{figure}
\begin{figure}[h!]
\centering
\includegraphics[width=14cm]{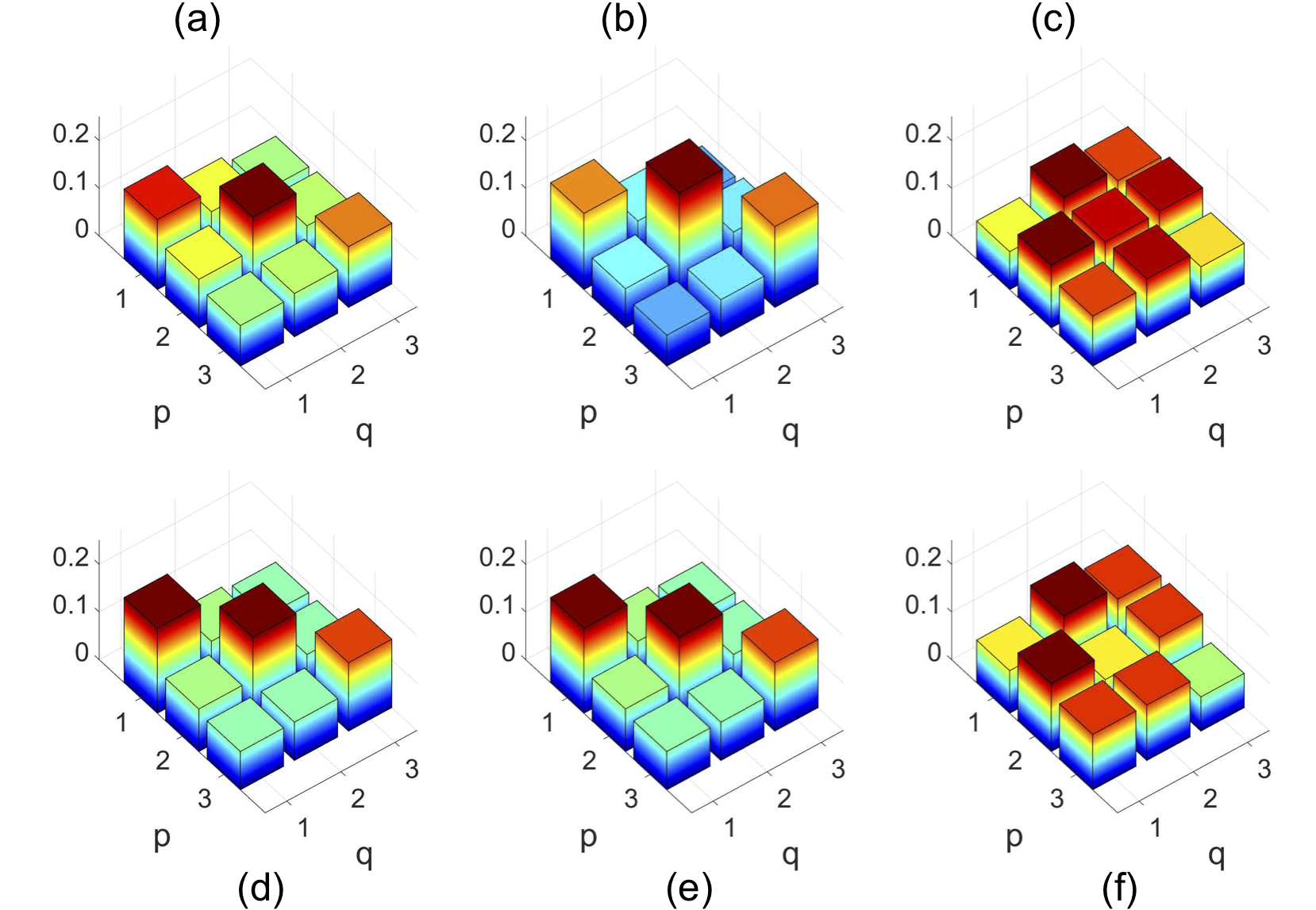}
\caption{Two-particle intensity correlations for separable (a, d), entangled (b, e), and incoherent photon pairs (c, e) after $z=12$ cm propagation. Upper row: Experimental data. Lower row: Numerical simulations.}
\label{fig:F22}
\end{figure}
\begin{figure}[!htb]
\centering
\includegraphics[scale=.5]{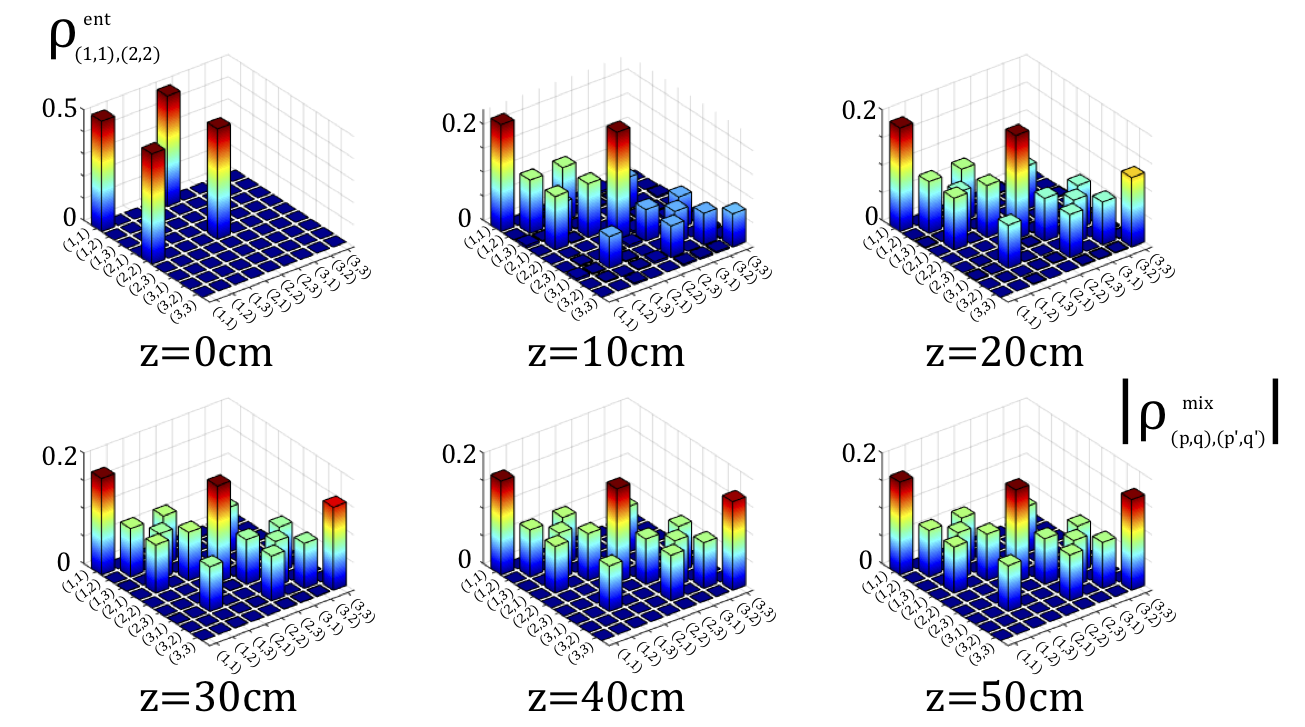}
\caption{Density matrices  (absolute value) for entangled two-photon states $\rho^{ent}_{(1,1),(2,2)}(0)$ propagating through waveguide trimers affected by a dephasing rate of $5\g_{exp}$.}
\label{fig:F7}
\end{figure}
\begin{figure}[!htb]
\centering
\includegraphics[scale=.5]{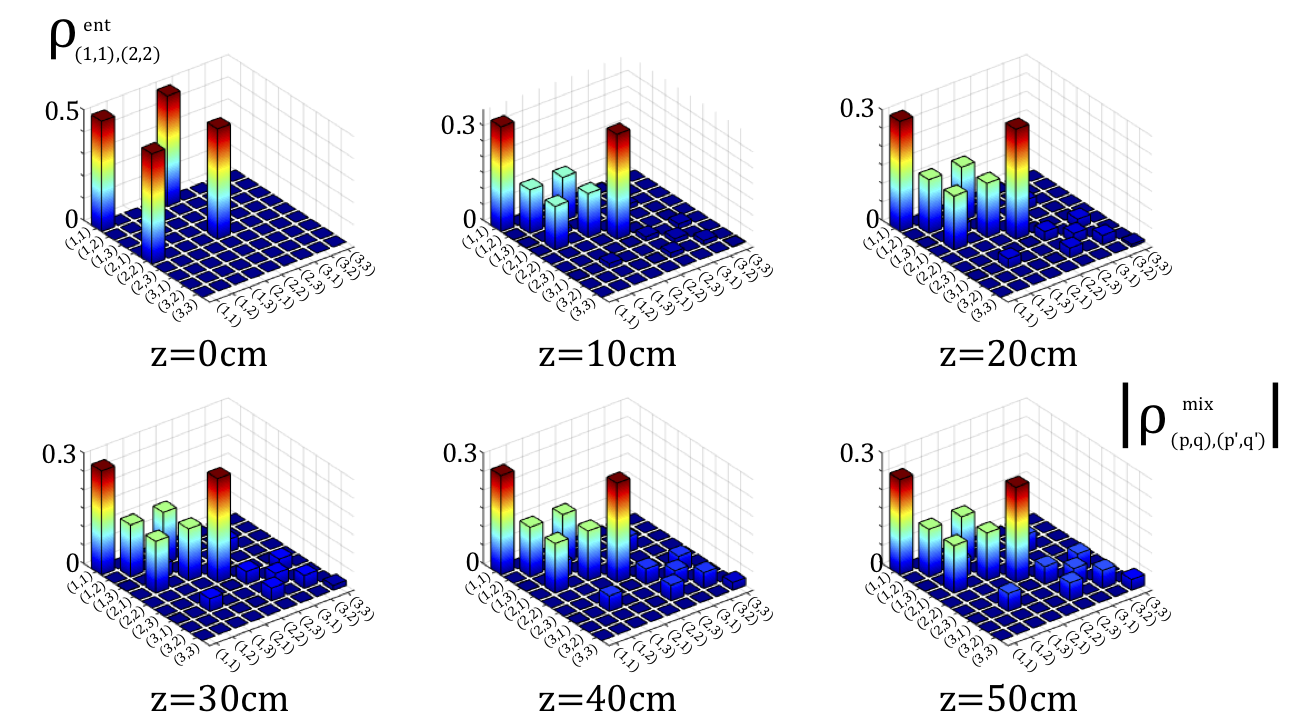}
\caption{Density matrices  (absolute value) for entangled two-photon states $\rho^{ent}_{(1,1),(2,2)}(0)$ propagating through waveguide trimers affected by a dephasing rate of $50\g_{exp}$.}
\label{fig:F9}
\end{figure}

\noindent
\textbf{INTRODUCTION}.\\  
The influence of random fluctuating environments over the evolution of dynamical systems has been a subject of intensive research since the beginning of modern science \cite{Brown1,Einstein2}. Particularly in quantum physics, environmental noise represents a prominent adversary that precludes the generation, control, and preservation of fundamental properties such as coherence, entanglement, and quantum correlations \cite{Breuer4,Vega,LoFranco,Benedetti1,Benedetti2}. Conventionally, quantum systems interacting with the environment are termed open quantum systems (OQS), and as such they constitute the most common structures encountered in nature. In this regard, the standard phenomenological approach to describe the evolution of OQS is the so-called Born-Markov approximation \cite{Breuer4}. In such an approach, the system-of-interest is weakly coupled to a large unstructured environment in such a way that the statistical properties of the latter remain unaffected \cite{Hoeppe6}. Generally, in the study of OQS the main interest is to explore the quantum properties exhibited by the system itself and not of the environment; as a result one restricts the investigation to the dynamics of the reduced-OQS only \cite{Breuer4,Alicki}. \\
In its simplest configuration, reduced-OQS subject to Born-Markov premises can be investigated in finite quantum networks in which environmental effects are modeled by introducing pure dephasing or, more generally, non-dissipative noise, into the site energies \cite{Eisfeld2013}.
At the single particle level, the relevance of such dephasing models have been highlighted in an interdisciplinary framework of studies ranging from biology \cite{Plenio10,Rebentrost11}, via quantum chemistry \cite{Park12,Caruso}, and electronics \cite{Leon-Montiel13} to photonics \cite{Biggerstaff15} and ultra-cold atoms \cite{Schonleber16}.  As it turns out, single-particle dephasing models do not show any divergence from wave mechanics \cite{Briggs1,Eisfeld2013,Briggs2,Briggs3,Torres18}. Rather the richness and complexity of genuine quantum processes are more prominent when a manifold of indistinguishable particles is considered \cite{Sansoni,DiGiuseppe,Franco,Beggi, Bose1, Omar, Paunkovic, Bellomo, Sciara}. \\
In any setting affected by dephasing, the phase properties of the associated quantum mechanical waves are randomly distorted, as a result their interference capability or `coherence' tends to vanish \cite{Streltsov,Baumgratz,Winter}.
Indeed, the fragility of quantum coherence is one of the main impediments for the development of quantum-enhanced technologies \cite{Streltsov}. Clearly, identifying mechanisms to prevent or slow down decoherence effects in quantum systems is an issue of scientific and practical importance. In this respect, it has been demonstrated, theoretically \cite{Bromley} and experimentally \cite{Silva},  that some types of quantum correlations can be arrested into judiciously prepared quantum states that resist the impact of non-dissipative noise. More precisely, it has been recognized that when multiqubit Bell-diagonal states are exposed to dephasing, quantum coherence defined with respect to certain reference basis remains unaffected indefinitely \cite{Silva}, that is, the `amount' of quantum coherence becomes stationary. 
Beyond this particular framework, there exist certain special sets of states, termed decoherence-free subspaces, that exhibit some sort of symmetry that gives rise to a common coupling with the environment \cite{Lidar}. Consequently, along evolution the entire subspace will undergo collective decoherence \cite{Lidar} that can be factored out from the process leaving the state undistorted \cite{Kwiat}.
\\
In the present work we investigate, theoretically and experimentally, the dynamics of single- and two-  non-interacting particles in quantum networks affected by non-dissipative noise, that is, networks subject to the Born-Markov approximation. In particular, we focus our attention on the theoretical description of the role of particle indistinguishability in the preservation of  coherence of two-particle states, bosons and fermions, propagating under the influence of dephasing. We stress that the choice of using non-dissipative noise for our analysis is based on the fact that the number of particles has to be preserved.
\\
To perform our experiments, we exploit the fact that the dynamics of true reduced-OQS can be effectively reproduced in ensembles of subsystems where the fluctuating parameters, either site energies or coupling parameters, can be implemented physically \cite{Chen}. Notice, the simulation of OQS with Hamiltonian ensembles is only possible for systems under the influence of non-dissipative noise. For our experiments, such reduced-like OQS are implemented in ensembles of photonic waveguide networks inscribed in fused silica glass by means of the femtosecond laser writing technique \cite{Szameit22}. In this photonic context, time is mapped onto the propagation coordinate $(t\rightarrow z)$, the waveguide propagation constants play the role of site energies, and the hopping rates result from the evanescent overlap between normal modes supported by adjacent waveguides (sites) \cite{Szameit22}. We emphasise that within the single excitation manifold the mapping of the time parameter over the propagation coordinate allows one to literarilly observe the evolution of transition amplitudes with time \cite{Weimann}. Further, dephasing effects can be produced in the waveguides by inducing random longitudinal fluctuations in the waveguides' refractive indices as illustrated in Fig.~(\ref{fig:F1} a).\vspace{1cm}\\
\noindent
\textbf{RESULTS}\\ 
\textbf{Single particle dynamics}\\
In order to produce stochastic fluctuations in the waveguide `site-energies', we have varied the inscription velocity while writing the waveguides at intervals of one centimeter. This effectively produces random changes in the site-energies every $\sim$33ps.
Mathematically, single-particle dynamics in such photonic networks is governed by the stochastic Schr\"odinger equation $-\mathrm{i}\frac{d}{dz} \psi_{n}(z)=\beta_{n}(z) \psi_{n}(z) + \sum_{m\ne n}^{N}\cc_{m,n}\psi_{m}(z).$ Here, we have set $\hbar=1$, $\p_{n}$ represents the single-particle wavefunction at site $n$, and $\cc_{m,n}$ are the hopping rates between the $(m,n)$ sites. Moreover, $\beta_{n}(z)=\beta_{n}+\phi_{n}(z)$ denotes the random site energy at the $n$-th site with $\phi_{n}(z)$ describing a stochastic Gaussian process which satisfies the conditions $\langle\phi_{n}(z)\rangle=0$ and $\langle\phi_{n}(z)\phi_{m}(z')\rangle=\g_{n}\delta_{m,n}\delta(z-z')$, with $\langle...\rangle$ denoting stochastic average, and $\g_{n}$ representing the dephasing rates \cite{Eisfeld2013}. Notice, the dephasing rates are directly obtained using the relation $\g_{n}=\sigma_{n}^{2}\triangle z$ \cite{Laing23}, where $\sigma_{n}$ is the standard deviation used to inscribe the $n$-th waveguide and $\triangle z=1$cm is the correlation length.
It is important to note that the assumption of uncorrelated Gaussian fluctuations in the site energies corresponds to the so-called Haken-Strobl model \cite{Rebentrost11}, which has been widely used to describe dephasing processes in realistic systems such as exciton transport in molecular aggregates and crystals, and photosynthetic complexes \cite{Eisfeld2013,Rebentrost11}.
Additionally, our experimental setup can be modified to reproduce more general, non-Gaussian or non-Markovian, stochastic processes. However, in such scenarios the number of particles is not necessarily preserved, as a result, we restrict our analysis to Gaussian noise.\\
In the presence of noise the proper instrument to describe quantum dynamical systems is the density matrix. Hence, by introducing the average single-particle correlation functions $\rho_{n,m}=\left\langle \p_{n}\p_{m}^{*}\right\rangle$ one obtains  the master equation for the reduced density matrix \cite{Eisfeld2013} (see Methods)
\begin{equation}\label{eq:m2}
\begin{aligned}
\im\frac{d}{dz}\rho_{n,m}&=\left[\left(\beta_{m}-\beta_{n}\right)-\frac{\im}{2}\left(\g_{n}+\g_{m}\right)\right] \rho_{n,m}+\im\sqrt{\g_{n}}\sqrt{\g_{m}} \delta_{n,m}\rho_{n,m}\\&-\sum_{r}\cc_{n,r} \rho_{r,m}+\sum_{r}\cc_{m,r}\rho_{n,r}.
\end{aligned}
\end{equation} 
Based on the fact that the optics of single-particles, bosons and fermions, is analogous to wave mechanics, here we experimentally analyze single-excitation dynamics utilizing laser light, see Fig.~(\ref{fig:F1} b) for a sketch of the experimental setup. Throughout this work we consider as demonstrative models waveguide trimers involving two relatively strongly-coupled sites both of which interact weakly with a third site, Fig.~(\ref{fig:F1} a). To fulfill these coupling requirements, we chose the coupling coefficients between the upper sites to be $\kappa_{1,2}=\kappa_{2,1}=2$cm$^{-1}$, while their coupling with the lower waveguide is $\kappa_{1,3}=\kappa_{2,3}=0.6$cm$^{-1}$. For all experiments the length of the samples was 12cm and the propagation constants were taken randomly from a Gaussian distribution with variance $\sigma=3$cm$^{-1}$ $\left(\sigma=2\right.$cm$\left.^{-1}\right)$ for the single-particle (two-particle) experiments, and mean values $\beta_{1}(z)=\beta_{2}(z)=1$cm$^{-1}$ and $\beta_{3}(z)=-1$cm$^{-1}$ for the upper and lower waveguides, respectively (see Methods).\\  
As reference case for the single-excitation manifold, we depict in Fig.~(\ref{fig:F1} c) the experimental intensity evolution of light traversing a noiseless trimer. Evidently, the light propagates in a coherent fashion hopping predominantly between the strongly-coupled sites (upper waveguides), and at most 10\% of the total energy hops into the farthest site (lower waveguide). In contrast, when the trimers become disturbed by noise, the regular hopping of the wavefunctions is no longer sustained. Instead, the average wave-packets evolve into an incoherent superposition of delocalized states. These effects are demonstrated experimentally by injecting light into one of the upper sites of an ensemble containing 21 different dynamically disordered trimers. Then, after averaging the intensities over the ensemble we find the pattern displayed in Fig.~(\ref{fig:F1} d). Notice at the propagation distance of $z=12$cm our observations reveal an homogeneous intensity distribution covering all sites. Interestingly, such homogeneousness in the intensity occurs despite the fact the associated waveguides are inscribed at different separation distances.\\ 
Concurrently, theoretical inspection of the off-diagonal elements $\rho_{n,m}$ elucidate a gradual decay of coherence. These effects are shown in Fig.~(\ref{fig:F2}) for different dephasing strengths. Notice in all cases the coherences $\rho_{n,m}$ inherently decay demonstrating that dephasing transforms single-particle states into stationary states with nullified coherence \cite{Plenio10,Rebentrost11,Caruso}. These results can be explained from the fact that the off-diagonal elements $\rho_{n,m}$ exhibit a complex \emph{propagation constant} $\left(\beta_{m}-\beta_{n}\right)-\im\left(\g_{n}+\g_{m}\right)/2$, where the negative imaginary part implies attenuation. On the contrary, for the diagonal elements, $\rho_{n,n}$, such propagation constants turn out to be zero.\\ 
These theoretical findings along with our experimental observations unequivocally confirm that single-excitations, subject to non-dissipative noise, coherently evolve during a certain time, and eventually the system reaches a steady state constituted of a uniform incoherent mixture of states \cite{Plenio10,Rebentrost11,Caruso}. For the sake of completeness, in Figs.~(\ref{fig:F1} e, f) we compare the theoretically computed diagonal elements $\rho_{n,n}=\left\langle \p_{n}\p_{n}^{*}\right\rangle$, and the experimental intensity distributions for the coherent and the dephasing case, respectively. The good agreement between the experimental and numerical results suggest that Eq. (\ref{eq:m2})  can be assumed valid.
\vspace{1cm}
\\
\textbf{Two-particle dynamics}\\  
In stark contrast to single-excitations, when two indistinguishable particles co-propagate in the same type of structures, interesting effects occur revealing that some of the corresponding coherence terms resist the impact of dephasing. Our theoretical framework is based on the concept of two-particle state $\PP_{p,q}$, which describes jointly two particles populating sites $(p,q)$ \cite{Abouraddy25}. In terms of $\PP_{p,q}$ we define the two-particle density matrix $\rho_{(p,q),(p',q')}=\left\langle\PP_{(p,q)}\PP_{(p',q')}^{*}\right\rangle$, whose diagonal elements, $\rho_{(p,q),(p,q)}=\left\langle\PP_{(p,q)}\PP_{(p,q)}^{*}\right\rangle=\left\langle|\PP_{(p,q)}|^{2}\right\rangle$, yield the joint probability density $G_{p,q}^{(2)}=\rho_{(p,q),(p,q)}$, also termed coincidence rate \cite{Gilead27}. 
It is important to note that there have been some previous studies investigating the impact of static disorder on the dynamics of two-particle intensity correlations, which represent the diagonal elements of the corresponding density matrix \cite{DiGiuseppe,Gilead27}.\\
In the presence of dynamic disorder, the two-particle master equation governing the system is given as (see Methods)
\begin{equation}\label{eq:TwoME}
\begin{aligned}
\im\frac{d}{dz}\rho_{(p,q),(p',q')}=\left[\left(-\beta_{p}-\beta_{q}+\beta_{p'}+\beta_{q'}\right)-\frac{\im}{2}(\g_{p}+\g_{q}+\g_{p'}+\g_{q'})\right]\rho_{(p,q),(p',q')}\\
+\im\left[ \sqrt{\g_{p}\g_{p'}}\delta_{p,p'}+\sqrt{\g_{p}\g_{q'}}\delta_{p,q'}
+\sqrt{\g_{q}\g_{p'}}\delta_{q,p'}+\sqrt{\g_{q}\g_{q'}}\delta_{q,q'}-\sqrt{\g_{p}\g_{q}}\delta_{p,q}-\sqrt{\g_{p'}\g_{q'}}\delta_{p',q'}\right]\rho_{(p,q),(p',q')}\\
- \sum_{r} \left[\cc_{r,p}\rho_{(r,q),(p',q')}+\cc_{r,q}\rho_{(p,r),(p',q')}-\cc_{r,p'}\rho_{(p,q),(r,q')}-\cc_{r,q'}\rho_{(p,q),(p',r)}\right].
\end{aligned}
\end{equation}
For our analysis, we consider the situations where the stochastic trimers are excited by two indistinguishable particles in pure separable states $\ket{\PP^{sep}}=\frac{1}{\sqrt{2}}\left(\ket{1_{1},1_{2}}\pm\ket{1_{2},1_{1}}\right)$, where the $\pm$ signs determine whether the particles are bosons $(+)$ or fermions $(-)$. It is worth noting that fermionic-like statistics with bosons is nowadays possible using the polarization degree of freedom of photons, e.g. \cite{Sansoni,Crespi}. 
The density matrices corresponding to these initial states, $\rho_{(1,2),(2,1)}^{bos}=\left(\ket{1_{1},1_{2}}\bra{1_{1},1_{2}}+\ket{1_{1},1_{2}}\bra{1_{2},1_{1}}+H.C.\right)/2$ and $\rho_{(1,2),(2,1)}^{fer}=\left(\ket{1_{1},1_{2}}\bra{1_{1},1_{2}}-\ket{1_{1},1_{2}}\bra{1_{2},1_{1}}+H.C.\right)/2$, are shown in Figs.~(\ref{fig:F3} a, h). For bosons we also examine the evolution of entangled two-particle states, 
$\ket{\PP^{ent}}=~\frac{1}{\sqrt{2}}\left(\ket{1_{1},1_{1}}+\ket{1_{2},1_{2}}\right)$, whose density matrix is depicted in Fig.~(\ref{fig:F3} b). Notice, throughout this work we use the compact notation $\ket{1_{m},1_{n}}$ to represent states where one particle is populating site $m$ and another site $n$, $\ket{1_{m}}\otimes\ket{1_{n}}$, while states $\propto\left(\ket{1_{m},1_{n}}+\ket{1_{n},1_{m}}\right)$, are symmetrized wavefunctions. In this convention the two-particle state $\PP_{m,n}$ corresponds to $\ket{1_{m},1_{n}}$. It is noteworthy that the off-diagonal terms present in the initial density matrices, Figs.~(\ref{fig:F3} a, h), arise by virtue of the wavefunction symmetrization needed to account for the indistinguishability and exchange statistics of the particles \cite{Franco,Sperling}. In other words, such off-diagonal elements appear due to the identicalness of the particles and in their present form it is not possible to identify two-particle coherence (entanglement) \cite{Franco}. It is also important to remark that quantum theory demands that pure states involving identical particles have to be described by (anti-)symmetrized state vectors. Otherwise, theoretical predictions may exhibit dramatic discrepancies with experimental observations, e.g. \cite{Herbut, McKeown}. We emphasize that the (anti-)symmetrization is referred to the particle statistics (bosons or fermions), which is the meaningful physical property, and not to the structure of the state with respect to the fictitious labels introduced to distinguish the particles.\\ 
Integration of the two-particle master equation Eq.~\eqref{eq:TwoME} with the initial states $\ket{\PP^{sep}}$ and $\ket{\PP^{ent}}$
renders the density matrices displayed in Fig.~(\ref{fig:F3}). These results clearly show that after a propagation distance of about 12 cm, the density matrices for separable and entangled bosons become identical, and after 20 cm the systems reach the steady state. Once in steady state, a closer inspection of the diagonal elements reveals that both particles bunch into the same site with probability $\rho_{(1,1),(1,1)}=\rho_{(2,2),(2,2)}=\rho_{(3,3),(3,3)}=0.15$, see Figs.~(\ref{fig:F3}~d). Concurrently, the remaining diagonal elements, those quantifying particle anti-bunching, exhibit the probabilities $\rho_{(1,2),(1,2)}=\rho_{(1,3),(1,3)}=\rho_{(2,1),(2,1)}=\rho_{(2,3),(2,3)}=\rho_{(3,1),(3,1)}=\rho_{(3,2),(3,2)}=0.09$. \\
Quite interestingly, our theory predicts the existence of some coherences in the resultant steady states as indicated by the off-diagonal elements exhibited in the density matrix, Figs.~(\ref{fig:F3}~d). To better appreciate these effects we decompose the steady state into four sub-matrices, $\rho_{bos}(z)=\rho_{(1,2),(2,1)}^{sep}+\rho_{(1,3),(3,1)}^{sep}+\rho_{(2,3),(3,2)}^{sep}+\rho_{(n,n),(n,n)}^{mix}$, where $\rho_{(n,n),(n,n)}^{mix}\propto \sum_{n=1}^{3}\ket{1_{n},1_{n}}\bra{1_{n},1_{n}}$ represents an incoherent superposition of two-particle states (classically-correlated two-particle states \cite{Abouraddy25}), and $\rho_{(p,q),(q,p)}^{sep}\propto \ket{1_{p},1_{q}}\bra{1_{p},1_{q}} + \ket{1_{p},1_{q}}\bra{1_{q},1_{p}}+H.C.$ is a coherent superposition of two-particle states. Indeed, the superposition $\rho_{(1,2),(2,1)}^{sep}+\rho_{(1,3),(3,1)}^{sep}+\rho_{(2,3),(3,2)}^{sep}$ implies that the particles coherently inhabit in all three sites with the same amplitude. Accordingly, within the steady state a mixture of both classically-correlated and coherent (indistinguishable) unentangled two-particle states perpetually coexist. Likewise, for indistinguishable fermion pairs the steady state exhibits some off-diagonal terms, see Fig.~(\ref{fig:F3}~j). However, unlike the bosonic case, the fermionic steady state is composed of only three sub-matrices $\rho_{fer}(z)=\rho_{(1,2),(2,1)}^{sep}+\rho_{(1,3),(3,1)}^{sep}+\rho_{(2,3),(3,2)}^{sep}$, thereby indicating the coexistence of quantum superpositions of indistinguishable two-fermion states. In fact, the absence of the matrix elements $\rho_{(n,n),(n,n)}^{mix}$ in the fermionic steady state is because according to the Pauli exclusion principle fermions cannot inhabit in the same site simultaneously.\\  
To explain the existence of coherence in the steady state, we see that the two-particle density matrix exhibits a complex \emph{propagation-constant} given by the first two terms on the right-hand-side of Eq.~\eqref{eq:TwoME}. For the diagonal elements $\rho_{(p,q),(p,q)}$ such \emph{propagation-constants} turn out to be zero, and the same occurs for the off-diagonal elements accounting for particle indistinguishability $\rho_{(p,q),(q,p)}$. The lack of the complex \emph{propagation-constant} in all diagonal and the off-diagonal elements $\rho_{(p,q),(q,p)}$ directly implies that they will remain immune to the impact of dephasing. 
Quantum mechanically, each particle from a pair of identical particles that travel through sites $p$ and $q$ affected by dephasing strengths $\gamma_{p}$ and $\gamma_{q}$, will experience the dephasing from both sites simultaneously. This implies that, due to the exchange symmetry,  the coherences due to particle indistinguishability will undergo the same `amount' of dephasing, as a result such coherences will remain undistorted. Indeed, this is an example of collective dephasing where the coherences are identically affected by the environment. Therefore, the subspace formed by the coherences $\rho_{(p,q),(q,p)}$ is a decoherence free subspace \cite{Lidar,Kwiat}.
Conversely, for the remaining off-diagonal elements $\rho_{(p,q),(p',q')}$ the \emph{propagation-constant} is given as $\left(-\beta_{p}-\beta_{q}+\beta_{p'}+\beta_{q'}\right)-\im\left(\g_{p}+\g_{q}+\g_{p'}+\g_{q'}\right)/2$. Owing to the negativity of the imaginary part, we determine that those elements will decay as they are affected by an attenuation factor arising from dephasing.
\\  
To shed light on the role of particle indistinguishability in the preservation of coherence, we consider the evolution of two-particle states exhibiting classical probabilities, $\rho_{(1,1),(2,2)}^{mix}=\left(\ket{1_{1},1_{1}}\bra{1_{1},1_{1}}+\ket{1_{2},1_{2}}\bra{1_{2},1_{2}}\right)/2$ and $\rho_{(1,2),(2,1)}^{inc}=\left(\ket{1_{1},1_{2}}\bra{1_{1},1_{2}}+\ket{1_{2},1_{1}}\bra{1_{2},1_{1}}\right)/2$. Physically, $\rho_{(1,1),(2,2)}^{mix}$ involves two indistinguishable particles entering together into either one of the upper sites of the trimer with exactly the same classical probability. That is, $\rho_{(1,1),(2,2)}^{mix}$ is a two-particle state presenting the strongest possible classical correlation \cite{Abouraddy25}. Conversely, the state $\rho_{(1,2),(2,1)}^{inc}$ is made of identical particles that are initially distinguishable.\\
Remarkably, for the initial state $\rho_{(1,1),(2,2)}^{mix}$, integration of the master equation, Eq.~\eqref{eq:TwoME}, renders a steady state which is identical to the ones obtained for separable and entangled bosons, Fig.~(\ref{fig:F3}~d). In contrast, $\rho_{(1,2),(2,1)}^{inc}$ yields a different density matrix where the anti-bunching elements, $\left(\rho_{(1,2),(1,2)}=\rho_{(1,3),(1,3)}=\rho_{(2,1),(2,1)}=\rho_{(2,3),(2,3)}\right.$ $\left.=\rho_{(3,1),(3,1)}=0.15\right)$, become larger than the bunching ones, $\rho_{(1,1),(1,1)}=\rho_{(2,2),(2,2)}=\rho_{(3,3),(3,3)}=0.09$. Interestingly, even though the initial state involves two distinguishable bosons, we identify some signatures of coherence in the density matrix, see off-diagonal elements in Fig.~(\ref{fig:F3}~g).  
The explanation behind the appearance of those coherences is that during evolution, the initially distinguishable particles can tunnel into the same site, and since the particles are identical when they reach the same site they become indistinguishable. Hence, a soon as the coherences manifest themselves in the state, our theory predicts that they will remain immune to dephasing.\\
In order to quantify the difference between all steady states we use the trace-distance criterion $D\left(\rho_{m},\rho_{n}\right)=\frac{1}{2}Tr|\rho_{m}-\rho_{n}|$, which yields zero if and only if $\rho_{m}=\rho_{n}$ \cite{Smirne}. In the present case $\rho_{m,n}$ represents any combination of steady states arising from the initial states $\left(\rho_{(1,2),(2,1)}^{bos},\rho_{(1,1),(2,2)}^{ent},\rho_{(1,1),(2,2)}^{mix},\rho_{(1,2),(1,2)}^{inc}\right)$. By doing so we find $D\left(\rho_{m},\rho_{n}\right)=0$ for all cases where the particles were indistinguishable since the begining, and $D\left(\rho_{m},\rho_{n}\right)=0.25$ for the cases comprising steady states of initially indistinguishable and distinguishable particles.\\
To quantify the amount of surviving coherence in the steady states, we use two different coherence measures, the physically intuitive norm of coherence $C_{n}(\rho)=\sum_{i\neq j}|\rho_{i,j}|$ \cite{Baumgratz}, and the relative entropy of coherence $C_{RE}(\rho)=S(\rho_{diag})-S(\rho)$ \cite{Streltsov}, with $S$ representing the von Neumann entropy and $\rho_{diag}$ the matrix obtained from the density matrices $\rho$ after removing all off-diagonal elements. Notice, in both measures, a totally mixed (incoherent) state is signaled by a vanishing coherence measure.
In Fig.~(\ref{fig:entropy}) we show the evolution of the norm of coherence and the entropy of coherence for the initial states separable (solid line), path-entangled (dashed line), and incoherent (dash-dotted line) states, respectively. Importantly, when distinguishable particles are injected into the system $(C_{n}=C_{RE}=0)$, coherence due to indistinguishability rapidly emerges and the system evolves into a steady-state where $C_{n}=0.25$ and $C_{RE}=0.06$.
From these theoretical results we convincingly state that under the influence of dephasing, identical particles always evolve into a steady state in which coherences due to indistingushability perpetually prevail. In the Methods section we present the resulting two-particle steady states for different dephasing rates, and we demonstrate that the same steady state occurs even in the presence of strong dephasing, see Figs.~(\ref{fig:F7}, \ref{fig:F9}). The extension of the theory to the case of $N$ particles should be straightforward as indicated in the Methods.\\
\noindent
As we pointed out above, the diagonal elements of the density matrices are equal to the joint particle probability density, $G_{p,q}^{(2)}=\rho_{(p,q),(p,q)}$. Hence, to experimentally prove the validity of the two-particle master equation, Eq.~\eqref{eq:TwoME}, we have performed two-photon intensity correlation measurements for separable, entangled, classically correlated, and distinguishable two-photon (two-boson) states using an ensemble of 37 waveguide trimers. The details regarding the state preparation and waveguide fabrication are given in the Methods section.\\
The experimental average photon coincidence measurements are depicted in Figs.~(\ref{fig:F4} e-g) where it is clear that, under the influence of dephasing, initial states involving indistinguishable photons are driven to undergo identical correlation patterns. More specifically, for all three cases involving indistinguishable two-particle states (separable, entangled, and classically correlated), the measurements reveal the tendency of the photons to bunch into the same site including the farthest weakly-coupled waveguide. Concurrently, photon coincidences (antibunching) occur with similar probabilities, but less frequently than bunching events as illustrated by the off-diagonal elements in Figs.~(\ref{fig:F4} e-g). Finally, when exciting the same stochastic networks with distinguishable (incoherent) photons coupled separately into the upper sites, the correlation patterns were found to exhibit the higher probabilities in the off-diagonals (antibunching) terms, Fig.~(\ref{fig:F4} h). Note the correlation matrices, $G_{p,q}^{(2)}=\rho_{(p,q),(p,q)}$, have been arranged in a way that the bunching elements, 
$\left(\rho_{(1,1),(1,1)},\rho_{(2,2),(2,2)},\rho_{(3,3),(3,3)}\right)$, are shown along the diagonal, while the anti-bunching terms, 
$\left(\rho_{(1,2),(1,2)},\rho_{(1,3),(1,3)},\rho_{(2,1),(2,1)},\rho_{(2,3),(2,3)},\rho_{(3,1),(3,1)},\rho_{(3,2),(3,2)}\right)$, are displayed in the off-diagonals entries.\\ 
Since the elements of the experimental correlation matrices represent the diagonal elements of the density matrix, we can compute the average fidelity between the experimental $G_{p,q}^{(2)-exp}$ and theoretical $G_{p,q}^{(2)-th}=\rho_{(p,q),(p,q)}$  two-particle probability densities. This is done using the expression $S=\left(\sum_{p,q}\sqrt{G_{p,q}^{(2)-exp} G_{p,q}^{(2)-th}}\right)^{2}/\sum_{p,q}G_{p,q}^{(2)-exp}\sum_{p,q} G_{p,q}^{(2)-th}$  \cite{peruzzo2010,Lebugle29}, which gives fidelities of $0.9935$, $0.9945$, and $0.9948$ for the entangled, separable, and incoherent input states, respectively.
In the Methods section we compare the experimental correlations, Fig.~(\ref{fig:F4} e-h), versus the diagonal elements of the theoretically-computed density matrices displayed in Fig.~(\ref{fig:F3}~d,~g), see Fig.~(\ref{fig:F22}).
\vspace{0.5cm}\\
\noindent
\textbf{DISCUSSION}.\\  
At this point it is worth emphasizing that there is an ongoing debate regarding the observability or physical significance of correlations due to symmetrization \cite{Benatti,Franco,Reusch}. Indeed, the controversy arises because correlations of the type $\left(\rho_{(p,q),(q,p)},\rho_{(q,p),(p,q)}\right)$ represent superpositions of two-particle states where the only difference is the order of the particles. Formally, such coherences do not represent manipulable superpositions, but their presence in the steady state implies that the particles retain their capability to interfere (indistinguishability) in experiments of the Hong-Ou-Mandel type \cite{Killoran30,Herbut,Zou,Salehx}. Indeed, it has been shown that superpositions originated from the exchange symmetry can become accessible and exploitable \cite{Cavalcanti,Killoran30,Roos2017, LoFAx1, LoFAx2}.\\
We stress the results presented in this work differ from the ones reported in references \cite{Bromley,Silva}. In the present case, we have shown that indistinguishable particles always evolve towards a steady state exhibiting some coherences that imply particle indistinguishability. On the other hand, in references \cite{Bromley, Silva} it is demonstrated that under certain conditions the amount of coherence exhibited by certain Bell-diagonal states remains stationary. Regarding our experimental setup, it differs from the one shown in reference \cite{Biggerstaff15}, in the fact that here we implement the effects of dephasing by modifying the properties of the waveguides. On the contrary, in  \cite{Biggerstaff15} the environmental effects are modeled by tuning the input field and the analysis is restricted to the single-excitation manifold.\\
In this work, we have investigated, theoretically and experimentally, Born-Markov OQS within the single and two-excitation manifolds. We showed that even when individual particles do not preserve any quantum coherence in the presence of noise, indistinguishable two-particle states retain, on average, quantum coherence despite the impact of dephasing. More importantly, we have shown that due to the exchange symmetry the prevailing coherences undergo collective dephasing, as a result, the set of coherences due to particle indistinguishability form a decoherence-free subspace.  
Finally, we point out that quantum technologies are frequently based on identical particles, bosons or fermions. That is, identical particles constitute the elementary building blocks of quantum systems. Hence, one can in principle encode information in the correlations due to particle indistinguishability, and since those correlations are robust against dephasing, the information can be transmitted through dephasing channels without any distortion.
\newpage
\clearpage
\newpage
\clearpage
\newpage
\clearpage
\newpage
\clearpage
\newpage
\clearpage
\newpage
\noindent
\textbf{METHODS}\\
\textbf{Devices fabrication and specifications}\\
The waveguides samples employed in the experiments were inscribed in high-purity fused silica (Corning 7980, ArF grade) using a RegA 9000 seeded by a Mira Ti:Al2O3 femtosecond laser. Pulses centred at 800nm with duration of 150fs were used at a repetition rate of 100 kHz with energy of 450nJ. The pulses were focused 250 $\mu$m under the sample surface using a numerical aperture = 0.6 objective while the sample was translated at different speeds ranging from 60 to 240 mm min$^{-1}$ for the classical experiments and from 70 to 131 mm min$^{-1}$ for the quantum. The samples were translated by a high-precision positioning stages (ALS130, Aerotech Inc.). 
The random changes in the speed inscription were chosen from a gaussian distribution at intervals of one centimeter, effectively producing stochastic fluctuations in the site-energies every $\sim$33ps. Note we have used two different sigmas because for the single-particle experiments the operation wavelength is 633 nm, while for two photons is 808nm. Hence, the parameters have been adjusted to fabricate waveguides to operate at the desired wavelengths.
The mode field diameters of the guided mode were 18$\times$20$\mu$m  at 815 nm. At the wavelength of interest, propagation losses and birefringence were estimated at 0.3 dB cm$^{-1}$ and in the order of 10$^{-7}$, respectively. The waveguides are equally spaced by 127$\mu$m at the input-output facets to couple into standard V-groove fibre arrays for the in- and out-coupling of the photons. The waveguides then smoothly converge through fanning arrangements to their eventual configuration in the functional sections.\vspace{1cm}\\
\textbf{Experiments with classical light}\\
To experimentally demonstrate the functionality of the suggested waveguide system, we use an ensemble of 21 different dynamically disordered waveguide trimers. The input signal is prepared by focusing a beam from a HeNe laser onto the front facet of the sample. Then, by exploiting the fluorescence from colour centres within the waveguides, we monitor the full intensity evolution from the input to the output plane by using a CCD camera. After recording the intensity from the 21 samples we perform the average intensity and the final result is presented in Fig.~(\ref{fig:F1} d).
The average of the fluorescence images, Fig.~(\ref{fig:F1} d), clearly shows the emergence of a uniform redistribution of energy among all the waveguides.\\
\textbf{Preparation of two-photon states}\\  
Photon pairs were generated at a wavelength $\lambda$=815nm using a standard type-I spontaneous-parametric-down-conversion source by pumping a BiBO crystal with a $407.5$nm continuous-wave laser diode at $70$mW. We employed commercial V-groove fiber arrays to couple the photons into the on-chip trimers as well as to collect them at the output facet of individual waveguides. We use high-NA multimode fibers to feed the photons into the respective avalanche photodiode. This in turn ensures low coupling losses at the output of the chip. From the data of the photodiodes, the photon probability distribution at the output, as well as the two-point intensity correlations, can be extracted.\\ 
To prepare indistinguishable separable photon pairs from a SPDC source we additionally apply filters with 3nm bandwidth as shown in Fig. (\ref{fig:F1} b). Entangled two-photon states were readily created by simultaneously exciting the two input modes of an integrated 50:50 directional coupler with indistinguishable photons in a separable product state \cite{Lebugle29}. Classically-correlated two-photon states were produced in a similar fashion with the difference that we induce a delay of $\sim$2ps in one of the output ports of the integrated 50:50 directional coupler such that we have two distinguishable two-photon states (classically-correlated states). Distinguishable two-photon pairs were produced by delaying one of the photons $\sim$2ps with respect to the other before entering the samples (this time without an additional integrated 50:50 directional coupler). We point out that a time delay of $\sim$2ps is sufficient to make the photons distinguishable and the distinguishability was verified by the absence of interference in a standard Hong-Ou-Mandel setup.\vspace{1cm}\\  
\textbf{Derivation of the single- and two-particle master equations}\\  
\textbf{Single-Excitation Manifold}\\  
We start by considering a stochastic network containing $N$ coupled sites.
In such configurations, the propagation of single-particle probability amplitudes are governed by the stochastic Schr\"odinger equation \cite{Eisfeld2013}
\begin{equation}\label{eq:1}
-\mathrm{i}\frac{d}{dz} \psi_{n}(z)=\beta_{n}(z) \psi_{n}(z) + \sum_{m\ne n}^{N}\cc_{m,n}\psi_{m}(z).
\end{equation}
Here we have set $\hbar=1$, $\psi_{n}(z)$ is the probability amplitude for a single-particle propagating through site $n$, $\beta_{n}(z)$ are the stochastic site energies which depend on the propagation distance $z$, and $\cc_{m,n}$ represents the coupling coefficients between sites $m$ and $n$.
In order to account for environmental effects, we assume random site energies varying according to the functions $\beta_{n}(z)=\beta_{n}+\phi_{n}(z)$, where $\phi_{n}(z)$ describes a stochastic Gaussian process satisfying the conditions
\begin{subequations}
\begin{equation}\label{eq:2a}
 \begin{aligned}
\langle \phi_{n}(z)\rangle&=0,
 \end{aligned}
\end{equation}
\begin{equation}\label{eq:2b}
 \begin{aligned}
\langle \phi_{n}(z) \phi_{m}(z')\rangle&=\gamma_{n}\delta_{m,n}\delta(z-z'),
 \end{aligned}
\end{equation}
\end{subequations}
with $\langle...\rangle$ denoting stochastic average. Note we have assumed the simplest scenario in which the system is affected by white noise, described by Eq.~\eqref{eq:2b}, where $\g_{n}$ denotes the noise intensity, $\delta_{n,m}$ is the Kronecker delta used to indicate that each site energy fluctuates independently from each other, and $\delta(z-z')$ is a Dirac delta describing the Markovian approximation \cite{vanKampen1981}.\\
Writing Eq.~\eqref{eq:1} in differential form we have
\begin{equation}\label{eq:3}
 \begin{aligned}
d\psi_{n}=\mathrm{i}\beta_{n}\psi_{n}dz+\mathrm{i}\sum_{r}\cc_{n,r}\psi_{r}dz+\mathrm{i}\psi_{n}\phi_{n}(z)dz,
 \end{aligned}
\end{equation}
and by introducing the so-called Wiener increments \cite{Laing23}
\begin{equation}\label{eq:4}
 \begin{aligned}
dW_{n} &= \frac{\phi_{n}(z)}{\sqrt{\gamma_{n}}}dz,\\
\bigl< dW_{n}dW_{m}\bigr>&=\delta_{n,m}dz,
 \end{aligned}
\end{equation}
we can cast Eq.~\eqref{eq:3} as
\begin{equation}\label{eq:5}
\begin{aligned}
d\p_{n}&=\im \beta_{n}\p_{n}dz+\im\sum_{r}\cc_{n,r}\p_{r}dz+\im\p_{n}\sqrt{\g_{n}}dW_{n}.
\end{aligned}
\end{equation}
We note that Eq.~\eqref{eq:5} has the so-called Stratonovich form \cite{vanKampen1981}. In order to compute the differential of the product $\p_{n}(z)\p_{m}^{*}(z)$, we use It\^{o}'s product rule $d(\p_{m}\p_{n}^{*})=d(\p_{n})\p_{m}^{*}+\p_{n}d(\p_{m}^{*})+d(\p_{n})d(\p_{m}^{*})$ \cite{vanKampen1981}, which demands $d\p_{n}$ to be written in It\^{o}'s form \cite{Eisfeld2013}
\begin{equation}\label{eq:6}
\begin{aligned}
d\p_{n}&=\left(\im \beta_{n}\p_{n}+\im \sum_{r}\cc_{n,r}\p_{r}-\frac{1}{2}\g_{n}\p_{n}\right)dz+\im \sqrt{\g_{n}}\p_{n}dW_{n}.
\end{aligned}
\end{equation}
Hence, using Eq.~\eqref{eq:6} we obtain the expression
\begin{equation}\label{eq:7}
\begin{aligned}
d(\p_{n}\p_{m}^{*})=\left[\im\left(\beta_{n}-\beta_{m}\right)-\frac{1}{2}\left(\g_{n}+\g_{m}\right)\right]\p_{n}\p_{m}^{*}dz+\im\sum_{r}\cc_{n,r}\p_{r}\p_{m}^{*}dz-\im\sum_{r}\cc_{m,r}\p_{n}\p_{r}^{*}dz\\+\im\sqrt{\g_{n}}\p_{n}\p_{m}^{*}dW_{n}-i\sqrt{\g_{m}}\p_{n}\p_{m}^{*}dW_{m}
+\sqrt{\g_{n}}\sqrt{\g_{m}}\p_{n}\p_{m}^{*}dW_{n}dW_{m},	
\end{aligned}
\end{equation}
where we have only considered terms up to first order in $dz$. Finally, by taking the stochastic average of Eq.~\eqref{eq:7} and defining the density matrix as $\rho_{n,m}=\left\langle\p_{n}\p_{m}^{*}\right\rangle$ we arrive to the evolution equation for the single-particle density matrix presented in the main text, Eq.~\eqref{eq:m2}.\\

\noindent
\textbf{Two-Excitation Manifold}\\
We now follow a similar procedure described above to derive the evolution equation for two-particle density matrices in coupled networks affected by dephasing. To do so, we start by considering pure two-particle states at sites $p$ and $q$ within a network comprising $N$ sites
\begin{equation}\label{eq:9}
\begin{aligned}
\PP_{p,q}(z)=\sum_{m,n}^{N,N}\varphi_{m,n}\left[U_{p,n}(z)U_{q,m}(z)\pm U_{p,m}(z)U_{q,n}(z)\right],
\end{aligned}
\end{equation}
where $\varphi_{m,n}$ is the initial probability amplitude profile $\left(\sum_{n,m}|\varphi_{m,n}|^{2}=1\right)$, and $U_{r,s}(z)$ represents the impulse response of the system, that is, the unitary probability amplitude for a particle traveling into site $r$ when it was initialized at site $s$. Moreover, the sign $+$ and $-$ determine whether the particles are bosons or fermions.\\
From Eq.~\eqref{eq:9} we define the product $\PP_{p,q}\PP_{p',q'}^{*}$, and using the It\^{o}'s product rule \cite{vanKampen1981} we compute the $z$-derivative as follows
\begin{equation}\label{eq:10}
\begin{aligned}
\frac{d}{dz}\left[\PP_{p,q}\PP_{p',q'}^{*}\right]=\left[\frac{d}{dz}\PP_{p,q}\right]\PP_{p',q'}^{*}+ \PP_{p,q}\left[\frac{d}{dz}\PP_{p',q'}^{*}\right]+\left[\frac{d}{dz}\PP_{p,q}\right]\left[\frac{d}{dz}\PP_{p',q'}^{*}\right].
\end{aligned}
\end{equation}
To compute Eq.~\eqref{eq:10} we need the It\^{o}'s form for the differential $\left[d\PP_{p,q}\right]$, which is given by
\begin{equation}\label{eq:11}
\begin{aligned}
d\PP_{p,q}=\im \left(\beta_{p}+\beta_{q}\right)dz \PP_{p,q} +\im \sum_{r} \left(\cc_{r,p} \PP_{r,q}+\cc_{r,q} \PP_{p,r}\right)dz -\frac{1}{2}(\g_{p}+\g_{q})\PP_{p,q}dz\\
+\im\left( \sqrt{\g_{p}}dW_{p}+ \sqrt{\g_{q}}dW_{q}\right) \PP_{p,q}-\sqrt{\g_{p}\g_{q}}\PP_{p,q}dW_{p}dW_{q}.
\end{aligned}
\end{equation}
Eq.~\eqref{eq:11} can be easily obtained by taking the derivative of Eq.~\eqref{eq:9} and using the fact that $U_{r,s}$ are single-particle probability amplitudes which obey Eq.~\eqref{eq:5}, namely
\begin{equation}\label{eq:12}
\begin{aligned}
dU_{p,n}&=\im \beta_{p}U_{p,n}dz+\im\sum_{r}\cc_{p,r}U_{r,n}dz+\im U_{p,n}\sqrt{\g_{p}}dW_{p}.
\end{aligned}
\end{equation}
Then, by defining the density matrix $\rho_{(p,q),(p',q')}=\left\langle\PP_{p,q}\PP_{p',q'}^{*}\right\rangle$, and after some algebra we obtain the  evolution equation for the two-particle density matrix presented in the main text, Eq.~\eqref{eq:TwoME}.\\
The generalization to $N$ indistinguishable particles is straightforward following similar steps as for the two-particle case and introducing the $N$-particle probability amplitude 
\begin{equation}\label{eq:14}
\begin{aligned}
\PP_{p,q,r,...}(z)=\sum_{a,b,c,...}^{N}\varphi_{a,b,c,...}\left[\chi_{a,b,c,...}^{p,q,r,...}+\chi_{a,b,c,...}^{per}+...\right],
\end{aligned}
\end{equation}
where we have defined $\chi_{a,b,c,...}^{p,q,r,...}=U_{p,a}(z)U_{q,b}(z)U_{r,c}(z)...$, with $U_{m,n}$ representing  the probability amplitude for a single-particle at site $n$ when it was launched at site $m$, and the superscript $per$ means cyclic permutations of superscripts $p$, $q$, $r$, $...$.\\    
In order to integrate Eq.~\eqref{eq:m2} and Eq.~\eqref{eq:TwoME} it is necessary to obtain the individual dephasing rates $\g_{m}$ for $m = 1, 2, 3$. This is easily done using the relation $\g_{m}=\sigma_{m}^{2}\triangle z$ \cite{Laing23,Montiel2014}, where $\sigma_{m}$ is the standard deviation of the $m$-th site, $\triangle z$ is the correlation length. To perform the simulations shown in Figs.~(\ref{fig:F2}) and (\ref{fig:F3}), we have used the $\sigma_{m}$ obtained from the data utilized to inscribe the waveguides
\begin{equation}\label{eq:15}
\begin{split}
\mathbf{\sigma_{exp}}=\left(\sigma_{1}=1.3143cm^{-1}, \sigma_{2}=1.3204cm^{-1}, \sigma_{3}=1.3283cm^{-1}\right), \quad Classical.\\
\mathbf{\sigma_{exp}}=\left(\sigma_{1}=1.1407cm^{-1}, \sigma_{2}=1.112cm^{-1}, \sigma_{3}=1.1371cm^{-1}\right), \quad Quantum.
\end{split}
\end{equation}
In both cases, classical and quantum, $\triangle z=1cm$. Hence, using these $\mathbf{\sigma_{exp}}$  we obtain the individual dephasing rates for numerical integration of Eq.~\eqref{eq:m2} and Eq.~\eqref{eq:TwoME}
 \begin{equation}\label{eq:16a}
 \begin{split}
\mathbf{\g_{exp}}=\left(\g_{1}=1.7275cm^{-1},\g_{2}=1.7435cm^{-1},\g_{3}=1.7645cm^{-1}\right), \quad Classical.\\
\mathbf{\g_{exp}}=\left(\g_{1}=1.3012cm^{-1},\g_{2}=1.2365cm^{-1},\g_{3}=1.2930cm^{-1}\right), \quad Quantum.
\end{split}
\end{equation}
\newpage
\noindent
\textbf{Experimental-theoretical Comparison of two-photon correlations}\\ 
In Fig.~(\ref{fig:F22}) we show the intensity correlation matrices for separable Figs.~(\ref{fig:F22} a, d), entangled Figs.~(\ref{fig:F22} b, e), and distinguishable (incoherent) photon pairs Figs.~(\ref{fig:F22} c, f) after a propagation distance of $z=12\text{cm}$. The upper row depicts the experimental coincidence measurements, whereas the lower one shows our theoretical predictions. \\
\newpage
\noindent
\textbf{Impact of strong dephasing on two-photon density matrices}\\ 
In order to elucidate the impact of dephasing over two-boson states coupled into the upper sites of the waveguide trimer shown in Fig~(\ref{fig:F1} a), we consider an entangled two-photon state 
\begin{equation}\label{eq:17}
\begin{aligned}
\rho^{ent}_{(1,1),(2,2)}(0)&=\frac{1}{2}\left(\ket{1_{1},1_{1}}\bra{1_{1},1_{1}}+\ket{1_{1},1_{1}}\bra{1_{2},1_{2}}+\ket{1_{2},1_{2}}\bra{1_{1},1_{1}}+\ket{1_{2},1_{2}}\bra{1_{2},1_{2}}\right),
\end{aligned}
\end{equation}
and perform numerical integration of Eq.~\eqref{eq:TwoME} for two different dephasing rates. Specifically, we employ dephasing rates $\gamma = 5\gamma_{exp}$, and $50\gamma_{exp}$, where $\gamma_{exp}$ represents the dephasing rate utilized in our experiments. These dephasing values correspond to changing proportionally the variance of the Gaussian distribution used to chose the random energies of the waveguides. For a dephasing rate of $\g=5\g_{exp}$, integration of Eq.~\eqref{eq:TwoME} renders the density matrices shown in Fig.~(\ref{fig:F7}). These results indicate that for the initial state $\rho^{ent}_{(1,1),(2,2)}(0)$, the steady state emerges at $z\approx40$cm. That is, the system reaches the steady state at approximately twice the distance with respect to the case when $\g=\g_{exp}$. For the second case where the dephasing rate is increased to $50\g_{exp}$, Fig.~(\ref{fig:F9}) indicates that the evolution towards the steady state becomes slower in comparison with the case $\g=5\g_{exp}$.
From these results we can state that in the presence of non-dissipative noise, the system will evolve towards the steady state either much slower or much faster depending on the dephasing strength: weak-to-moderate dephasing will drive the system into its steady state faster than strong dephasing will do. At first sight, these results seem to be counter-intuitive, however, the effects of high dephasing can be thought
of as an example of the quantum Zeno effect where the evolution of the system is suppressed by a rapid dephasing. That is, the Zeno effect will suppress the transport and keep the excitation localized in the two upper sites.
\newpage
\newpage
\noindent
Data availability\\
Source code and data are available from the authors upon reasonable request.
\section*{ACKNOWLEDGEMENTS}
\noindent
R.J.L.M. acknowledges financial support by CONACYT-M\'exico project CB-2016-01/284372, and DGAPA-UNAM project UNAM-PAPIIT IA100718. 
Armando Perez-Leija, Kurt Busch, and Alexander Szameit acknowledge financial support by the Deutsche Forschungsgemeinschaft (PE 2602/2-2,
BU 1107/12-2, BU 1107/10-1 and SZ 276/9-2, SZ 276/12-1).
\section*{AUTHOR CONTRIBUTIONS}
APL, DGS, and RJLM contributed equally to this work. A.P.L. and R.J.L.M. initiated the project and developed the theory and simulations, D.G.S., M.G., and MH designed the samples, D.G.S. carried out the experiments, H.M.C, K.B. and A.S. supervised the project.
\section*{ADDITIONAL INFORMATION}
\section*{Competing interests:} The authors declare no competing financial interests.

\end{document}